\documentclass[structabstract]{aa}

\usepackage{amsmath}
\usepackage{graphicx}
\usepackage{txfonts}
\usepackage{natbib}

\bibpunct{(}{)}{;}{a}{}{,} 

\usepackage{multirow}
\newcommand\tstrut{\rule{0pt}{2.5ex}}
\newlength{\digit}
\newcommand{\dsp}{\hspace*{\digit}}

\newlength{\dotdigit}
\begin{document}

\title{Impact inducted surface heating by planetesimals on early Mars}

   \author{T.~I.~Maindl
          \inst{1}
          \and
          R.~Dvorak
          \inst{1}
          \and
          H.~Lammer
          \inst{2}
          \and
          M.~G\"{u}del
          \inst{1}
          \and
          C.~Sch\"afer
          \inst{3}
          \and
          R.~Speith
          \inst{4}
          \and
          P.~Odert
          \inst{5}
          N.~V.~Erkaev
          \inst{6,7}
          K.~G.~Kislyakova
          \inst{2}
          \and
          E.~Pilat-Lohinger
          \inst{5}}

\offprints{T.~I.~Maindl\\
\email{thomas.maindl@univie.ac.at}}

\institute{Department of Astrophysics, University of Vienna,  T\"{u}rkenschanzstrasse 17, A-1180 Wien, Austria
         \and
             Space Research Institute, Austrian Academy of Sciences, Schmiedlstrasse 6, A-8042 Graz, Austria\and
             Institut f\"ur Astronomie und Astrophysik, Eberhard Karls Universit\"at T\"ubingen, Auf der Morgenstelle 10, 72076 T\"ubingen, Germany
         \and
             Physikalisches Institut, Eberhard Karls Universit\"at T\"ubingen, Auf der Morgenstelle 14, 72076 T\"ubingen, Germany
         \and
             Institute of Physics, IGAM, University of Graz, Universit\"atsplatz 5, A-8010  Graz, Austria
          \and Institute for Computational Modelling, 660041 Krasnoyarsk 36, Russian Academy of Sciences, Russian Federation
          \and Siberian Federal University, 660041 Krasnoyarsk, Russian Federation}
\date{Received \today}

\abstract{}
{We investigate the influence of impacts of large planetesimals and small planetary embryos on the early Martian surface on the hydrodynamic escape of an early steam atmosphere that is exposed to the high soft X-ray and EUV flux of the young Sun.}
{Impact statistics in terms of number, masses, velocities, and angles of asteroid impacts onto the early Mars are determined via n-body integrations. Based on these statistics, smoothed particle hydrodynamics (SPH) simulations result in estimates of energy transfer into the planetary surface material and according surface heating. For the estimation of the atmospheric escape rates we applied a soft X-ray and EUV absorption model and a 1-D upper atmosphere hydrodynamic model to a magma ocean-related catastrophically outgassed steam atmosphere with surface pressure values of 52\,bar H$_2$O and 11\,bar CO$_2$.}
{The estimated impact rates and energy deposition onto an early Martian surface can account for substantial heating. The energy influx and conversion rate into
internal energy is most likely sufficient to keep a shallow magma ocean liquid for an extended period of time. Higher surface temperatures keep the outgassed steam
atmosphere longer in vapor form and therefore enhance its escape to space within $\sim$0.6\,Myr after its formation.}
{}


\maketitle

\section{Introduction}
Theoretical hypotheses based on geochemical observations indicate the occurrence
of magma oceans or at least magma ponds during the early evolution of terrestrial
planets but also in large planetary embryos and in many early accreting planetesimals
\citep{Elkins-Tanton-2012}. Impacts are a particular form of accretion. The melting during collisions between large planetesimals and planetary embryos suggests that silicate and metallic material may be processed through multiple magma oceans
before a growing planet reaches solidity. Impacts and the related
processes of magma ocean formation and its solidification, strongly influence
the earliest compositional differentiation, volatile contents and the origin of catastrophically outgassed H$_2$O and carbon-rich protoatmospheres of the terrestrial
planets \citep{Elkins-Tanton-2008, Elkins-Tanton-2012, Lammer-2013}.

Besides impacts, large planetesimals with radii between tens to hundreds of kilometers that
accreted within $\leq$1.5\,Myr most likely have experienced significant and in many cases
complete melting due to radiogenic heating from short-lived radioisotopes \citep{Urey-1955, Lee-et-al-1976, LaTourrette-Wasserburg-1998}.

Mars most likely formed before Venus and the Earth-Moon system \citep{Kleine-et-al-2004}.
Latest research in planet formation reveals
that Mars formed within a few Myr \citep{Brasser-2013, Morbidelli-et-al-2012}. Thus,
Mars can also be considered as a surviving large planetary embryo whose
building blocks consisted of material that formed in orbital locations just beyond the
ice line with an initial H$_2$O inventory of $\sim$0.1--0.2 wt-\%. \citet{Erkaev-et-al-2014}
showed that after the solidification of Mars' magma ocean, a catastrophically outgassed steam atmosphere
within the range of $\sim$50--250\,bar H$_2$O and $\sim$10--55\,bar CO$_2$ could have been
lost via hydrodynamic escape caused by the high EUV flux of the young Sun during $\sim$0.4--12\,Myr,
if the impact related energy flux of large
planetesimals and smaller planetary embryos to the planet's surface prevented
the steam atmosphere from condensing.

For Mars-size planetary embryos at 1.5\,AU, \citet{lebmas13}
studied the thermal evolution of early magma oceans in interaction with a catastrophically
outgassed steam atmosphere, and found that H$_2$O vapor would start to condense into liquid
water $\sim$0.1\,Myr after formation if one neglects frequent impacts by large planetesimals or smaller planetary embryos. However, such short condensation-time scales contradict
the isotopic analysis of Martian SNC meteorites \citep{debbra07},
where data can be best explained by a progressive crystallization of a magma
ocean with a duration of up to $\sim$100\,Myr. Because of this reason \citet{lebmas13}
suggested that frequent impacts of large planetesimals and small embryos, which have been neglected
in their study, may have kept the surface hotter during longer times. In such a case one will
obtain a hotter surface that
prevents atmospheric H$_2$O vapor from condensing \citep{Hayashi-et-al-1979, Genda-and-Abe-2005}.
For large planetary embryos that orbit in closer location around their host star, a steam atmosphere remains in vapor form much longer and is eventually lost to space without condensing to liquids.
Thus, the question whether impact induced surface heating keeps steam atmospheres
in vapor form longer or prevents the fast condensation of water vapor into liquid H$_2$O is more important for
planetary bodies such as Mars that orbit within the ice line but beyond Earth's orbit location.

Because impact inducted surface heating by planetesimals on early Mars and large Martian size planetary embryos
in general is crucial for the evolution and growth of terrestrial protoplanets and their initial volatile inventories
we investigate this process in detail. In Sect.~\ref{sect:impactheating} we discuss the impact statistics in the early
Solar System at Mars' orbit, the impact simulations and applied model, and finally the results.
Section~\ref{sect:implications} elaborates on the implications of our results to magma ocean-based catastrophically outgassed
steam atmospheres on early Mars and large planetary embryos in general. Section~\ref{sect:conclusion} concludes the study.

\section{Surface heating by impacts}
\label{sect:impactheating}
We investigate the influence of large-scale asteroid impacts on the surface temperature of the larger collision partner.
While the total energy involved in an asteroid impact is given by the kinetic energy of the impactor it is just an upper
limit for the energy that will be available for heating of the impact site material
\citep[see e.g., analytic estimations by][]{cel13}. We simulate impact events numerically via our own smoothed-particle hydrodynamics (SPH) code and track the efficiency of
converting kinetic energy of projectiles into inner energy in the impact region. Given the amount by which the inner
energy of the material involved in an impact process increases, we estimate the temperature rise in that area.

\subsection{Impact statistics}
\label{sect:impactstatistics}

To understand the duration of a magmatic ocean on early Mars or similar planetary embryos at orbital locations at 1.5\,AU, there exist several
recent studies, where the time scales estimated are quite different.
As mentioned before there is a discrepancy between theoretical models \citep[0.1\,Myr,][]{lebmas13} and observations \citep[100\,Myr after an geochemical analysis of SNC meteorites,][]{debbra07}. 
Explicitely, theoretical studies do not take into account the impacts of large planetesimals and their possible contribution
to the surface temperature on Mars in this early stage
during and just after the formation of the planets.

\begin{figure}
\begin{center}
\includegraphics[trim=100 70 40 80, clip, width=0.8\columnwidth, height=0.8\columnwidth
,angle=270
]{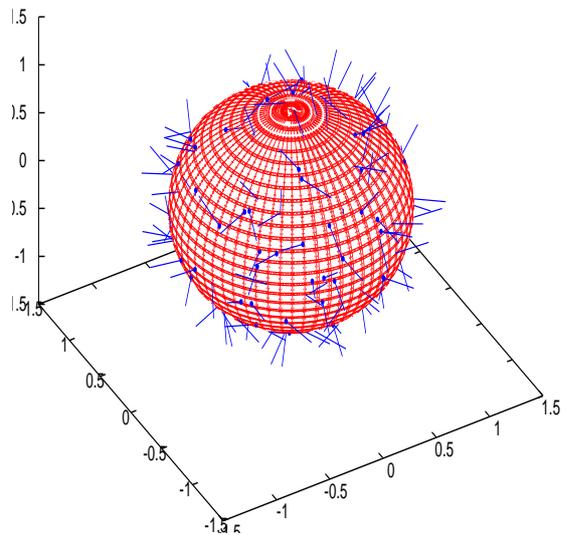}
\caption{Examples of impacts on Mars: the length of the lines are
proportional to the impact velocity, the small dots at the end of the lines
are the locations of the impacts on the surface. Note that the length does not differ much from one impact to the other (compare Fig.~\ref{fig:impvel} for the impact velocities).}
\label{fig:hedgehog}
\end{center}
\end{figure}
For determining the statistics of the impact velocities and impact angles on early
Mars during the young
phases of the Solar System we have undertaken extensive numerical
integrations in
different dynamical models \citep[cf.][]{maidvo13}.  In one model (MI) we distributed planetesimals of
different sizes in the region close to Mars with semimajor axes $a\/$ between $1.3\,\mathrm{AU} < a < 1.8\,\mathrm{AU}$
and eccentricities  $e < 0.15$; in another one (MII) we distributed them in the
region around $a\sim 3\,\mathrm{AU}\/$ with larger eccentricities ($e \sim 0.5$) and consequently
they have larger velocities in the distance of Mars (close to their
perihelion). In Fig.~\ref{fig:hedgehog}, we show a sample of impacts on the surface of Mars, all
from the MI model.

\begin{figure}
\begin{center}
\includegraphics[height=\columnwidth,angle=270,clip]{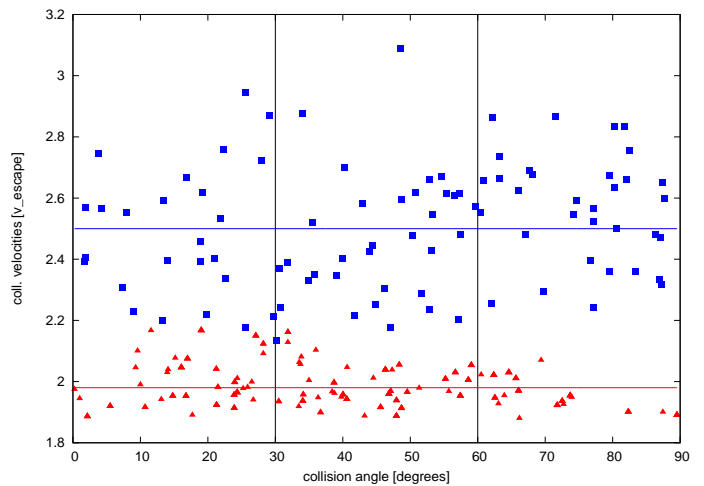}
\caption{Impact velocities (y-axis) versus the collision angle (x-axis) for a sample of our runs in MI (solid red triangles) and MII (solid blue squares).
The two horizontal lines show the mean values at $1.98\,v_\mathrm{esc}\/$ and $2.52\,v_\mathrm{esc}\/$ in
the respective models. The vertical lines divide the impact angles into three
equal intervals; note that whereas the impacts
in model MI are rare between 60 and 90 degrees, the impacts in MII are almost
equally distributed.}
\label{fig:impvel}
\end{center}
\end{figure}
\begin{figure}
\begin{center}
\includegraphics[height=\columnwidth,angle=270,clip]{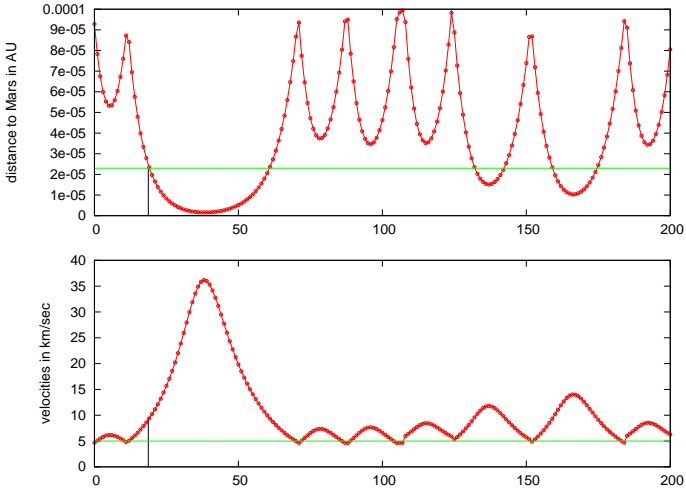}
\caption{Detailed scenario of the impact velocities: distance to the center of
 Mars (upper graph) and velocity (lower graph) versus
 time-integration steps during an encounter (x-axis). The horizontal lines represent the Martian radius and the surface escape velocity, respectively. The vertical lines mark the collision event. Note that the time steps taken by the adaptive step-size integration scheme (red dots) vary according to the acceleration of the body
between 0.02 days at a distance of $10^{-4}$\,AU to a few minutes close to the center.}
\label{fig:closeenc}
\end{center}
\end{figure}
It turned out that in model MII the collision velocities are only some $20\%$
larger than in the MI model; the latter showing impact velocities of about 10\,km s$^{-1}$ which is around twice the surface escape
velocity of Mars $v_\mathrm{esc}\/$. We depict the results in
Fig.~\ref{fig:impvel} where we plot the collision
velocities versus the impact angles for the two models. The blue squares
show collisions in MII, the red triangles in MI. The relatively large velocities
can be explained by the fact that inside the Hill
sphere (0.00386\,AU) the gravitation force of the planet is dominating the
Sun's gravitation and the respective planetocentric orbit is
a hyperbola with increasing velocities closer to the planet. In
Fig.~\ref{fig:closeenc} we show 8 close encounters inside a sphere of 0.0001\,AU around Mars. Out of
these encounters of a planetesimal with the planet three are `real'
collisions with Mars (the green line in the upper graph denotes the radius of
Mars at $2.3\cdot 10^{-5}\,\mathrm{AU}\/$). In our integration of the equations of motion with an
n-body code \citep[Lie-integration, see][]{handvo84}, where
we use mass points for all bodies involved, the orbit seems to continue
even inside Mars, but in fact we determine the impact velocity and impact
angle just at the surface of Mars. The upper limit of the
integration time was in
some cases 10\,Myr and the number of bodies was in the order of several
thousands with different masses. In MI we fully took into account
the gravitational
interaction of all bodies involved, in MII ($e \sim 0.5$) we used a simplified
model -- the elliptic restricted three body problem --, which was found to
give similar results compared to the full n-body problem for highly eccentric
orbits (but only for large eccentricities!).

To find out how many bodies of a certain mass collide with Mars during a
period of 1\,Myr we determined first of all the number of collisions of one
single body with Mars during the integration. For the determination of
the mass distribution in this stage
of the early Solar System around Mars we used the minimum-mass solar
nebula estimation with a surface density of
$\Sigma (a)= \Sigma_o a^{-\beta}\,\mathrm{g\, cm^{-2}}\/$ where $a\/$ is to be expressed in AU.
Different estimations for the parameters $\Sigma_o\/$ and $\beta\/$ were critically reviewed e.g.\ by \citet{kuc04}; instead of the older value of \citet{wei77} who used
$\Sigma_o = 4200$ we prefer to use the one by \citet{hay81}:
$\Sigma_o = 1700$, $\beta=1.5$. According to this formula we compute
the mass in a ring between $1\,\mathrm{AU} < a <
2\,\mathrm{AU}$ which is in the order of $80\, \mathrm{M_{Earth}}$. Following this distribution one
can see that the mass is almost equally
distributed in this range: from $9.8\, \mathrm{M_{Earth}}\;  (1.0 < a < 1.1)$\,AU
to $7.2\, \mathrm{M_{Earth}}\;  (1.9 < a < 2)$\,AU. These values were taken to estimate the
collisions of bodies in the two rings with small eccentricities and Mars.
In this range we now took a distribution of the sizes and masses according to a power law given in \citet{wya09}
which reads
\begin{align}
\sigma(D) \sim D^{2-3q}
\label{eq:sigmaD}
\end{align}
with $D\/$ denoting the body diameter. We consider body sizes from $D=200\/$\,m to $D=100\/$\,km. From this formula it follows that with $q\/$ values between 5/3 to 2 the number is dominated by small objects whereas the mass is dominated by the big bodies. We connected our estimates of the collisions from MI which turned out to be in the order of 1.5 collisions per Myr with this distribution, which leads to the energies released due to the impacts of different-size bodies presented in Table~\ref{t:impactstats}. For these impacts we just used the impact rates of MI because they are orders of
magnitudes larger than in MII.

\begin{table}
\centering
\caption[]{\label{t:impactstats}Impact statistics onto Mars in the early Solar System. The first column $N_\mathrm{i}\/$ gives the number of impacts per Myr, $r_\mathrm{P}=D/2\/$ the impactors' radii representative for the size intervals, and $\dot{E}_\mathrm{kin,all}$ the kinetic energy influx of all impactors of a certain size interval based on basalt asteroids with average density $\rho=2.7\,\mathrm{g\,cm^{-3}}\/$ and $10\,\mathrm{km\,s^{-1}}$ impact velocity. The values are obtained setting $q=2\/$ in (\ref{eq:sigmaD}) and a total mass of $80\, \mathrm{M_{Earth}}$.}
\begin{tabular}{lll}
\hline \hline
$N_\mathrm{i}\/$\tstrut & $r_\mathrm{P}$ & $\dot{E}_\mathrm{kin,all}$\\
{}[Myr$^{-1}$] & [km] & [$10^{27}$\,J\,Myr$^{-1}$]\\
\hline
$4.5$           &     50   & \dsp 0.318\tstrut \\
$4.5\cdot 10^1$ &     25   & \dsp 0.398 \\
$4.5\cdot 10^2$ &     15   & \dsp 0.859 \\
$4.5\cdot 10^3$ &     10   & \dsp 2.54 \\
$4.5\cdot 10^4$ & \dsp 5   & \dsp 3.18 \\
$4.5\cdot 10^5$ & \dsp 3   & \dsp 6.87 \\
$4.5\cdot 10^6$ & \dsp 1.5 & \dsp 8.59 \\
$4.5\cdot 10^7$ & \dsp 0.5 & \dsp 3.18 \\
$4.5\cdot 10^8$ & \dsp 0.1 & \dsp 0.254 \\
\hline
\tstrut & Total: & 26.19 \\
\hline
\end{tabular}
\end{table}

\subsection{Impact simulations}

\subsubsection{Impact modeling}
We model impacts onto the Mars surface with our 3-D solid-body continuum mechanics SPH code introduced and discussed in \citet{sch05} and \citet{maisch13} that includes the full elasto-plastic continuum mechanics model as formulated e.g., in \citet{maidvo13b} and implements the Grady-Kipp fragmentation model \citep{grakip80} for treating fracture and brittle failure as discussed in \citet{benasp94}. First-order consistency is achieved by applying a tensorial correction along the lines of \citet{schspe07}; dissipation of kinetic energy into heat is modeled via tracking inner energy including viscous energy terms originating from artificial viscosity \citep{mongin83}. As we expect the inner energy processes to happen immediately after the impact with negligible contributions from re-accreted ejecta, we do not include self gravity in the calculations.

The Mars surface and projectile material behavior is modeled via the Tillotson equation of state \citep{til62} assuming that both are made of basalt. The according parameters along with the Weibull distribution parameters are the same as used by \citet{maisch13}.

\begin{figure}
\centering
\includegraphics[width=0.85\columnwidth,clip]{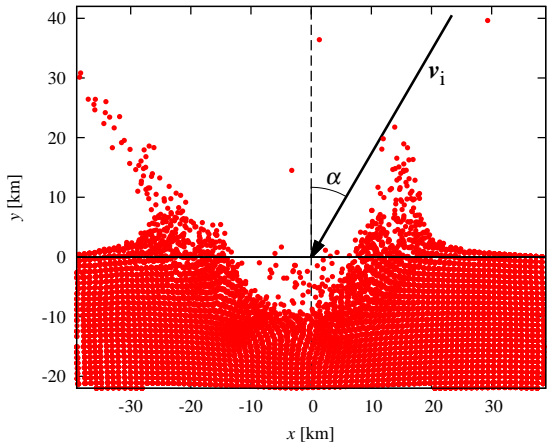}
\caption{\label{fig:geometry}Impact geometry -- the projectile hits the surface of Mars (horizontal line) at velocity $v_\mathrm{i}\/$ at an angle $\alpha\/$. The snapshot shows the situation 75 seconds after the impact of a 5\,km-diameter asteroid hitting the surface at $v_\mathrm{i}=7.5\,\mathrm{km\,s^{-1}}\/$ and $\alpha=30^\circ\/$. The data shown is a cut along the plane spawned by the projectile's velocity vector and a vector perpendicular to the planet's surface.}
\end{figure}
Our impact simulations use approx.\ 500,000 SPH particles per scenario and model part of the Mars surface as the target and a spherical projectile that impacts the target at an impact angle $\alpha\/$ and velocity $v_\mathrm{imp}\/$. The angle is defined such that a vertical ``head-on'' impact corresponds to $\alpha=0$ (cf.\ Fig.~\ref{fig:geometry}). Experiments with fewer SPH particles (100k, 200k, 300k, and 400k) confirm that the chosen resolution is suitable as the numerical values of the quantities of interest converge even for smaller particle numbers. The dimensions of the target are chosen such that boundary effects are minimized during the integration timespan (the kinetic and inner energy levels reach an equilibrium before the shock front reaches the boundary of the modeled volume). Due to keeping the actual number of SPH particles constant at $N_\mathrm{SPH}=489,285\/$ the (constant) smoothing lengths and the time interval $\delta t\/$ between output frames (the time integration itself always uses an adaptive step-size though) differ for different-size scenarios (characterized by the projectile radius $r_\mathrm{P}$): $\delta t=r_\mathrm{P}\cdot 2\cdot 10^{-4}\/$\,s, the smoothing lengths vary between 125\,m and 18.7\,km for the investigated $r_\mathrm{P}$-range of $100\,\mathrm{m}\le r_\mathrm{P}\le 15\,\mathrm{km}$.

\subsubsection{Simulation results}
\begin{figure}
\includegraphics[width=\columnwidth,clip]{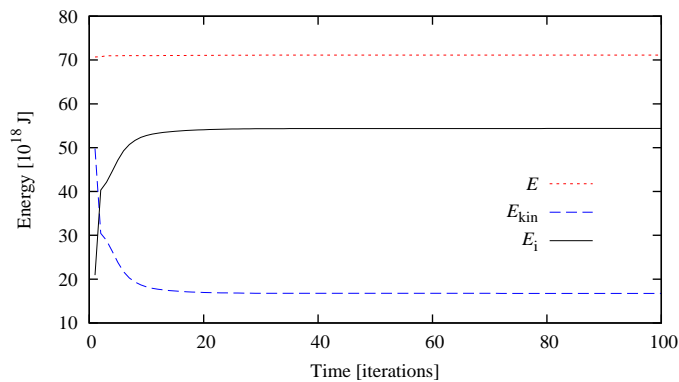}
\caption{Example for the time evolution of total energy $E\/$ (dotted red), kinetic energy $E_\mathrm{kin}\/$ (dashed blue), and inner energy $E_\mathrm{i}\/$ (solid black). This is the $r_\mathrm{P}=500\,\mathrm{m}\/$, $v_\mathrm{i}=10\,\mathrm{km\,s^{-1}}\/$, $\alpha=30^\circ\/$ scenario. One time iteration corresponds to 0.1\,s.}
\label{fig:ecurve}
\end{figure}

We find that as expected a large fraction of the available energy $E\/$ -- which equals the projectile's kinetic energy -- is
converted into inner energy $E_\mathrm{i}\/$ and is available for heating the surrounding matter. As shown in Fig.~\ref{fig:ecurve}
for an example, the inner energy is stationary very quickly after the impact (after less than 2\,s in the shown example of a
$r_\mathrm{P}=500\,\mathrm{m}\/$ impactor), which confirms our assumption of neglecting gravity in the simulations.

\begin{figure}
\includegraphics[width=\columnwidth,clip]{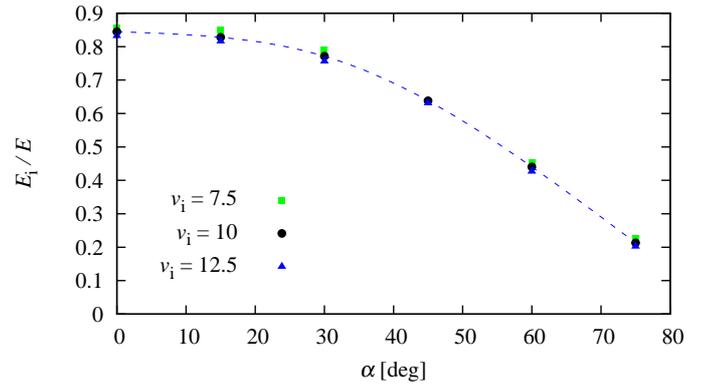}
\caption{Energy conversion efficiency for different impact angles. The fraction of the total available impact energy $E=E_\mathrm{kin,\,projectile}\/$ that is converted into inner energy $E_\mathrm{i}\/$ is largely independent from the impactor size and velocity $v_\mathrm{i}$ (only dependency on velocity shown), but significantly decreases for inclined impacts (large impact angle $\alpha\/$). The impact velocity $v_\mathrm{i}$ is given in $\mathrm{km\,s^{-1}}$.}
\label{fig:ei}
\end{figure}
While there is only a weak dependency of the stationary $E_\mathrm{i}/E\/$-values on the projectile size $r_\mathrm{P}\/$ (less than 0.01 over the investigated $r_\mathrm{P}\/$-interval) and the impact velocity $v_\mathrm{i}\/$, the converted kinetic energy decreases significantly for larger impact angles. Figure~\ref{fig:ei} shows the conversion efficiency for impacts with $v_\mathrm{i}=7.5$, 10, and 12.5\,$\mathrm{km\,s^{-1}}\/$ (equaling about 1.5, 2, and 2.5 $v_\mathrm{esc}\/$, respectively). The points correspond to average values of collision scenarios with projectiles between 100\,m and 15\,km radius -- the variations are about the size of the plotting symbols -- for impact angles between 0 and 75 degrees. While vertical impacts convert about 85\,\%, only slightly more than 20\,\% of the energy are available for heating at impacts at $\alpha = 75^\circ\/$. The remaining kinetic energy is consumed by ejecta and material displacement that is higher for inclined impacts.

\begin{figure}
\includegraphics[width=\columnwidth,clip]{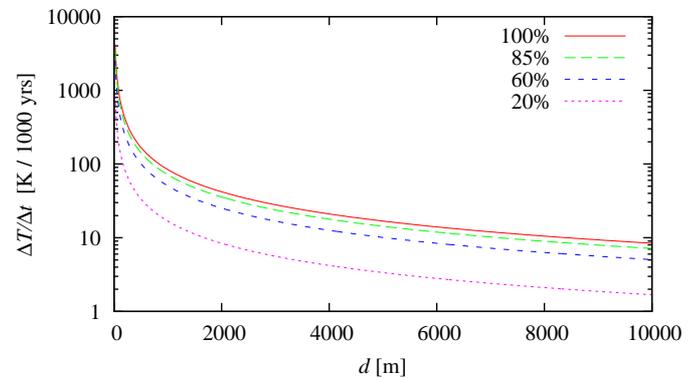}
\caption{First approximation to the temperature increase $\Delta T\/$ per 1000 years from inner energy deposited into the planetary surface for different conversion efficiencies $E_\mathrm{i}/E\/$ from 20\,\% to 100\,\%. It is assumed that thermal conductivity only takes place in a layer of thickness $d\/$.}
\label{fig:tplot}
\end{figure}
In principle the energy deposition $\Delta E\/$ onto the planetary surface can be converted to a temperature increase. Neglecting cooling processes and 
assuming a uniform bombardment the heating $\Delta T\/$ of a surface layer of depth $d\/$ can be expressed as
\begin{align}
\Delta T = \Delta E\,\left\lbrace  \frac{4\pi}{3}\,c_\mathrm{p}\,\rho\,\left[r_\mathrm{Mars}^3 - \left( r_\mathrm{Mars}-d\right)^3\right]\right\rbrace ^{-1}
\end{align}
with the surface material density $\rho\/$, specific heat $c_\mathrm{p}\/$ and Mars' mean radius $r_\mathrm{Mars}\/$. The parameter $d\/$ should be chosen to match the depth where heating processes from the planet's interior start to occur. Assuming an average specific heat capacity close to that of basalt ($c_\mathrm{p}=800\,\mathrm{J\,kg^{-1}\,K^{-1}}\/$), heat propagation only in a layer of thickness $d\/$, and adopting the total kinetic energy of the projectiles from Table~\ref{t:impactstats} a first approximation to the temperature increase for different conversion efficiencies $E_\mathrm{i}/E\/$ is given in Fig.~\ref{fig:tplot}. Considering the dominant MI collision scenario (cf.\ Sect.~\ref{sect:impactstatistics}) the average efficiency will be $\gtrsim 60\,\%\/$.

One should note that in our model for the conversion of $E_\mathrm{kin}\/$ into $E_\mathrm{i}\/$ we assumed the surface -- as well as the impactors -- to consist of solid basalt. While this holds for larger asteroids like those under consideration ($r_\mathrm{P}\gtrsim 100\,\mathrm{m}\/$), the liquid state of an existing magma ocean is beyond this model. As we have shown however, the heating takes place immediately after the impact and hence in the immediate neighborhood of the impact site where the material is completely damaged. From the simulation point of view it then behaves like a liquid and it is up to future investigations to study the effects of varying material parameters onto the energy conversion phenomena.

\section{Implications to catastrophically outgassed steam atmospheres}
\label{sect:implications}
According to \citet{lebmas13}, who studied the thermal evolution of an early
Martian magma ocean in interaction with a catastrophically outgassed $\sim$43\,bar
H$_2$O and $\sim$14\,bar CO$_2$ steam atmosphere, water vapor from such an atmosphere would start to condense at an orbit location of $\sim$1.5\,AU into liquid H$_2$O after $\sim$0.1\,Myr. For the estimation of the escape rates and thus the stability of such an atmosphere we apply the radiation absorption and a non-stationary 1-D hydrodynamic upper atmosphere model that
solves the hydrodynamic equations for mass, momentum and energy conversation
in spherical coordinates which
is described in detail in \citet{Erkaev-et-al-2013, Erkaev-et-al-2014} to the steam atmosphere assumed by \citet{lebmas13}. As described in \citet{Erkaev-et-al-2014} one can assume that dissociation products of H$_2$O molecules and CO$_2$ molecules should also
populate the lower hydrogen dominated thermosphere; we apply the same method, discussed in
detail in \citet{Hunten-et-al-1987}, \citet{Zahnle-et-al-1990} and \citet{Erkaev-et-al-2014}, to the loss of these heavier species that are dragged by the dynamically outward flowing bulk atmosphere.

The incoming high XUV flux \citep{Guedel-2007}, which heats the thermosphere, decreases due to absorption near the meso\-pause/homo\-pause level through dissociation and ionization of H$_2$O and H$_2$ molecules. We assume that atomic hydrogen is the dominant species in the upper atmosphere.
We estimate the heating efficiency that corresponds to the fraction
of absorbed XUV radiation which is transformed into thermal
energy to be 15\%. This value is in agreement with various studies
\citep{Chassefiere-1996, Yelle2004,Shematovich-et-al-2014}.

As in \citet{Erkaev-et-al-2014} and in agreement with \citet{Kasting-and-Pollack-1983} and \citet{Tian2005} we assume an atomic hydrogen density of $10^{13}$ cm$^{-3}$ at the lower boundary i.e., the meso\-pause/homo\-pause level of the hydrogen-rich upper atmosphere. According to \citet{Marcq2012} who used a 1-D radiative-convective atmospheric model to study the coupling between magma oceans
and overlaying steam atmospheres for corresponding surface temperatures that are
within a range of a few hundred to a few thousand Kelvin, the meso\-pause/homo\-pause level
can move to higher altitudes. As discussed in \citet{Erkaev-et-al-2014} in case of a
low gravity body such as Mars, this altitude where the atmospheric temperature is similar to the effective temperature, can -- for surface temperatures of
around 500--1000 K -- reach an altitude of $\sim$1000\,km above the planet's surface.
Because of the expansion of the steam atmosphere above the hot surface we apply our
model to a meso\-pause/homo\-pause location (i.e. lower boundary level) at 1000\,km.

The results of our impact study indicate that a frequent bombardment of large planetesimals within the size-range shown in Table~\ref{t:impactstats} could have contributed to a temperature enhancement of several hundred Kelvin near the surface so that the surface temperature rises to higher values than that of $\sim$500 K modeled for the so-called ``Mush'' stage by \citet{lebmas13}.
One should also note that for the surface temperatures of $\sim$500 K, which are expected during the ``Mush'' stage,
according to \citet{Kasting-1988} also H$_2$O vapor mixing ratios at the mesopause level will be $\sim$1. For that reason H$_2$O should continue
to escape effectively, even if there are periods of condensed liquid water on the planet's surface. However, frequent impacts as modeled in this study will also evaporate lakes or most likely prevent the formation of large lakes or oceans. On the other hand as discussed in \citet{Lammer-et-al-2013} impacts may also deliver additional volatiles 
that could be incorporated in the planetary environment after the XUV flux of the young Sun has decreased after the first few 100\,Myr so that a secondary atmosphere could grow.
However, the surface temperature enhancement by frequent impacts during the first 100\,Myr should keep a catastrophically outgassed steam atmosphere at 1.5\,AU in vapor form longer than 0.1\,Myr, at least periodically.

\begin{figure}
\centering
\includegraphics[width=0.9\columnwidth,clip]{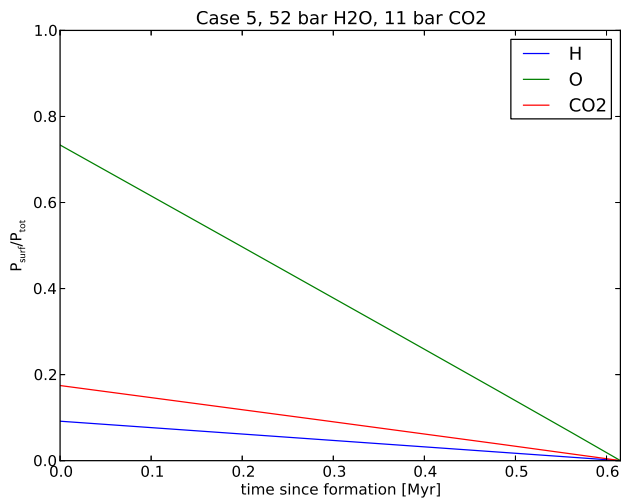}
\caption{Temporal evolution of the partial surface pressures $P_\mathrm{surf}$ of H, O, and CO$_2$ normalized to the total initial surface pressure $P_\mathrm{total}$ for an
catastrophically outgassed atmosphere as discussed in \citep{lebmas13}. The hydrogen inventory evolves assuming a constant escape rate of $\sim 7\times 10^{32}$\,s$^{-1}$
valid for 100 XUV. Both O and even CO$_2$ molecules are dragged along with the escaping H atoms.}
\label{fig:surfp}
\end{figure}

We assume the before mentioned initial conditions and the steam atmosphere as supposed by \citet{lebmas13} and expose it to a XUV flux during the activity saturation phase of the young Sun that was $\sim$44 times higher compared to that of the present Sun at 1.5\,AU \citep{Ribas-et-al-2005, Claire-et-al-2012}. Our escape model yields a H escape rate of $\sim 7\times 10^{32}$ s$^{-1}$ and the loss of an outgassed steam atmosphere with 52\,bar H$_2$O and 11\,bar CO$_2$ \citep{lebmas13,Erkaev-et-al-2014} after $\sim$0.6\,Myr. According to \citet{Erkaev-et-al-2014} such an initially outgassed steam
atmosphere corresponds to a magma ocean depth of about 500\,km and initial
water and CO$_2$ mixing ratios of 0.1\,wt-\% and 0.02\,wt-\%, respectively.
Denser steam atmospheres of up to about 260\,bar H$_2$O and 55\,bar CO$_2$
that may originate from deeper magma oceans and wetter building blocks
as studied by \citet{Erkaev-et-al-2014} will also be lost in agreement with their study within a period of 3\,Myr.
Figure~\ref{fig:surfp} shows the temporal evolution of the partial surface pressures $P_\mathrm{surf}$ of H, O, and CO$_2$ normalized to the total initial surface pressure $P_\mathrm{total}$ for an
catastrophically outgassed atmosphere of 52\,bar H$_2$O and 11\,bar CO$_2$. The hydrogen inventory evolves assuming a constant escape rate of $\sim 7\times 10^{32}$ s$^{-1}$. Both O and even CO$_2$ molecules are dragged along with the escaping H atoms.

Our finding agrees with
the model results of \citet{Erkaev-et-al-2014} and the hypothesis of \citet{Albarede-2007} that the initial catastrophically
outgassed Martian H$_2$O and volatile inventory was lost from the protoplanet within the
first 100\,Myr. The loss of the small planet's initial H$_2$O inventory
may also be a reason that early Mars did not develop a real plate tectonic regime because
of the fast hydrodynamic water loss. Water delivered by later impacts during the late veneer or late heavy bombardment phase \citep{Albarede-2009}, may have been again incorporated by hydrothermal alteration processes such as serpentinization, so that remaining parts of it could be stored in subsurface serpentine even today \citep{Chassefiere-et-al-2013}.

\section{Conclusion}
\label{sect:conclusion}
We showed that current estimates on impact rates and energies onto Mars in its early phase can account for substantial heating of its surface. Without a cooling mechanism the energy influx and conversion rate into internal energy is sufficient to keep a shallow magma ocean liquid for an extended period of time. This results
in higher surface temperatures that keep the outgassed steam atmosphere in vapor form for a longer time and enhances the escape to space.
We applied a XUV absorption and 1-D hydrodynamic upper atmosphere model to a catastrophically outgassed 52\,bar H$_2$O and CO$_2$
steam atmosphere to a 100 times higher XUV flux as expected during the first 100\,Myr of the young Sun. We find
that under such extreme solar activity-conditions such a steam atmosphere was most likely lost by hydrodynamic escape within $\sim$0.6\,Myr. The efficient escape of the bulk hydrogen gas drags heavier atoms and even CO$_2$ molecules to space.
Our results are also in agreement with \citet{Tian2009}, \citet{Lammer-et-al-2013} and \citet{Erkaev-et-al-2014} in
that early Mars most likely could not build up a dense CO$_2$ atmosphere during the early Noachian because of rapid escape
and impact induced surface heating. Depending on the impact frequency the results of our study also support the SNC-meteorite based hypothesis \citep{debbra07} according to which a shallow magma ocean or impact related sporadically produced shallow magma oceans could have existed during the first 100\,Myr on the Martian surface.
After the high activity of the young Sun and the impact frequency had decreased, a secondary atmosphere which has been outgassed by volcanic activity or has been delivered by smaller impactors $\sim$4--4.2 Gyr ago could have originated.

\begin{acknowledgements}
R.~Dvorak, M.~G\"{u}del, K.~G.~Kislyakova, H.~Lammer, T.~I.~Maindl, and E.~Pilat-Lohinger acknowledge the Austrian Science Fund (FWF) for supporting this study via FWF NFN project S11601-N16 ``Pathways to Habitability:
From Disks to Stars, Planets and Life'' and the related FWF NFN subprojects,
S11603-N16 ``Transport of Water and Organic Material in Early Planetary Systems'',
S11604-N16 ``Radiation \& Wind Evolution from T Tauri Phase to ZAMS and Beyond'',
S11607-N16 ``Particle/Radiative Interactions with Upper Atmospheres
of Planetary Bodies Under Extreme Stellar Conditions'', and S11608-N16 ``Binary Star Systems and Habitability''.
P.~Odert acknowledges support from the FWF project P22950-N16.
N.~V.~Erkaev acknowledges support by the RFBR grant No 12-05-00152-a.
This publication was supported by FWF.
The calculations for this work were in part performed on the hpc-bw-cluster -- we gratefully thank the bwGRiD project\footnote{\tiny bwGRiD (http://www.bw-grid.de), member of the German D-Grid initiative, funded by the Ministry for Education and Research (Bundesministerium fuer Bildung und Forschung) and the Ministry for Science, Research and Arts Baden-Wuerttemberg (Ministerium fuer Wissenschaft, Forschung und Kunst Baden-Wuerttemberg).} for the computational resources.
Finally, H.~Lammer thanks E.~Marcq from
LATMOS, Universit\'{e} de Versailles Saint-Quentin-en-Yvelines, Guyancourt, France
and L.~Elkins-Tanton from the Department of Terrestrial Magnetism, Carnegie Institution for Science, Washington DC 20015, USA,
for discussions related to magma oceans and outgassed steam atmospheres.
\end{acknowledgements}

\bibliographystyle{aa} 
\bibliography{references,lammer} 

\begin{thebibliography}{47}
\expandafter\ifx\csname natexlab\endcsname\relax\def\natexlab#1{#1}\fi

\bibitem[{{Albar{\`e}de}(2009)}]{Albarede-2009}
{Albar{\`e}de}, F. 2009, \nat, 461, 1227

\bibitem[{{Albar{\`e}de} \& {Blichert-Toft}(2007)}]{Albarede-2007}
{Albar{\`e}de}, F. \& {Blichert-Toft}, J. 2007, Comptes Rendus Geoscience, 339,
  917

\bibitem[{{Benz} \& {Asphaug}(1994)}]{benasp94}
{Benz}, W. \& {Asphaug}, E. 1994, \icarus, 107, 98

\bibitem[{{Brasser}(2013)}]{Brasser-2013}
{Brasser}, R. 2013, \ssr, 174, 11

\bibitem[{{Celebonovic}(2013)}]{cel13}
{Celebonovic}, V. 2013, \rmxaa, 49, 221

\bibitem[{{Chassefi{\`e}re}(1996)}]{Chassefiere-1996}
{Chassefi{\`e}re}, E. 1996, \jgr, 101, 26039

\bibitem[{{Chassefi{\`e}re} {et~al.}(2013){Chassefi{\`e}re}, {Langlais},
  {Quesnel}, \& {Leblanc}}]{Chassefiere-et-al-2013}
{Chassefi{\`e}re}, E., {Langlais}, B., {Quesnel}, Y., \& {Leblanc}, F. 2013,
  Journal of Geophysical Research (Planets), 118, 1123

\bibitem[{{Claire} {et~al.}(2012){Claire}, {Sheets}, {Cohen}, {Ribas},
  {Meadows}, \& {Catling}}]{Claire-et-al-2012}
{Claire}, M.~W., {Sheets}, J., {Cohen}, M., {et~al.} 2012, \apj, 757, 95

\bibitem[{{Debaille} {et~al.}(2007){Debaille}, {Brandon}, {Yin}, \&
  {Jacobsen}}]{debbra07}
{Debaille}, V., {Brandon}, A.~D., {Yin}, Q.~Z., \& {Jacobsen}, B. 2007, \nat,
  450, 525

\bibitem[{{Elkins-Tanton}(2008)}]{Elkins-Tanton-2008}
{Elkins-Tanton}, L.~T. 2008, Earth and Planetary Science Letters, 271, 181

\bibitem[{{Elkins-Tanton}(2012)}]{Elkins-Tanton-2012}
{Elkins-Tanton}, L.~T. 2012, Annual Review of Earth and Planetary Sciences, 40,
  113

\bibitem[{{Erkaev} {et~al.}(2014){Erkaev}, {Lammer}, {Elkins-Tanton},
  {St{\"o}kl}, {Odert}, {Marcq}, {Dorfi}, {Kislyakova}, {Kulikov},
  {Leitzinger}, \& {G{\"u}del}}]{Erkaev-et-al-2014}
{Erkaev}, N.~V., {Lammer}, H., {Elkins-Tanton}, L., {et~al.} 2014,
  http://dx.doi.org/10.1016/j.pss.2013.09.008

\bibitem[{{Erkaev} {et~al.}(2013){Erkaev}, {Lammer}, {Odert}, {Kulikov},
  {Kislyakova}, {Khodachenko}, {G{\"u}del}, {Hanslmeier}, \&
  {Biernat}}]{Erkaev-et-al-2013}
{Erkaev}, N.~V., {Lammer}, H., {Odert}, P., {et~al.} 2013, Astrobiology, 13,
  1011

\bibitem[{{Genda} \& {Abe}(2005)}]{Genda-and-Abe-2005}
{Genda}, H. \& {Abe}, Y. 2005, \nat, 433, 842

\bibitem[{{Grady} \& {Kipp}(1980)}]{grakip80}
{Grady}, D.~E. \& {Kipp}, M.~E. 1980, {International Journal of Rock Mechanics
  and Mining Sciences \& Geomechanics Abstracts}, 17, 147

\bibitem[{{G{\"u}del}(2007)}]{Guedel-2007}
{G{\"u}del}, M. 2007, Living Reviews in Solar Physics, 4, 3

\bibitem[{{Hanslmeier} \& {Dvorak}(1984)}]{handvo84}
{Hanslmeier}, A. \& {Dvorak}, R. 1984, \aap, 132, 203

\bibitem[{{Hayashi}(1981)}]{hay81}
{Hayashi}, C. 1981, Progress of Theoretical Physics Supplement, 70, 35

\bibitem[{{Hayashi} {et~al.}(1979){Hayashi}, {Nakazawa}, \&
  {Mizuno}}]{Hayashi-et-al-1979}
{Hayashi}, C., {Nakazawa}, K., \& {Mizuno}, H. 1979, Earth and Planetary
  Science Letters, 43, 22

\bibitem[{{Hunten} {et~al.}(1987){Hunten}, {Pepin}, \&
  {Walker}}]{Hunten-et-al-1987}
{Hunten}, D.~M., {Pepin}, R.~O., \& {Walker}, J.~C.~G. 1987, \icarus, 69, 532

\bibitem[{{Kasting}(1988)}]{Kasting-1988}
{Kasting}, J.~F. 1988, \icarus, 74, 472

\bibitem[{{Kasting} \& {Pollack}(1983)}]{Kasting-and-Pollack-1983}
{Kasting}, J.~F. \& {Pollack}, J.~B. 1983, \icarus, 53, 479

\bibitem[{{Kleine} {et~al.}(2004){Kleine}, {Mezger}, {M{\"u}nker}, {Palme}, \&
  {Bischoff}}]{Kleine-et-al-2004}
{Kleine}, T., {Mezger}, K., {M{\"u}nker}, C., {Palme}, H., \& {Bischoff}, A.
  2004, \gca, 68, 2935

\bibitem[{{Kuchner}(2004)}]{kuc04}
{Kuchner}, M.~J. 2004, \apj, 612, 1147

\bibitem[{{Lammer}(2013)}]{Lammer-2013}
{Lammer}, H. 2013, {Origin and Evolution of Planetary Atmospheres}

\bibitem[{{Lammer} {et~al.}(2013){Lammer}, {Chassefi{\`e}re}, {Karatekin},
  {Morschhauser}, {Niles}, {Mousis}, {Odert}, {M{\"o}stl}, {Breuer}, {Dehant},
  {Grott}, {Gr{\"o}ller}, {Hauber}, \& {Pham}}]{Lammer-et-al-2013}
{Lammer}, H., {Chassefi{\`e}re}, E., {Karatekin}, {\"O}., {et~al.} 2013, \ssr,
  174, 113

\bibitem[{{LaTourrette} \& {Wasserburg}(1998)}]{LaTourrette-Wasserburg-1998}
{LaTourrette}, T. \& {Wasserburg}, G.~J. 1998, Earth and Planetary Science
  Letters, 158, 91

\bibitem[{{Lebrun} {et~al.}(2013){Lebrun}, {Massol}, {Chassefi{\`e}Re},
  {Davaille}, {Marcq}, {Sarda}, {Leblanc}, \& {Brandeis}}]{lebmas13}
{Lebrun}, T., {Massol}, H., {Chassefi{\`e}Re}, E., {et~al.} 2013, Journal of
  Geophysical Research (Planets), 118, 1155

\bibitem[{{Lee} {et~al.}(1976){Lee}, {Papanastassiou}, \&
  {Wasserburg}}]{Lee-et-al-1976}
{Lee}, T., {Papanastassiou}, D.~A., \& {Wasserburg}, G.~J. 1976, \grl, 3, 109

\bibitem[{{Maindl} \& {Dvorak}(2014)}]{maidvo13}
{Maindl}, T.~I. \& {Dvorak}, R. 2014, in IAU Symposium, Vol. 299, IAU
  Symposium, ed. M.~{Booth}, B.~C. {Matthews}, \& J.~R. {Graham}, 370--373

\bibitem[{{Maindl} {et~al.}(2014){Maindl}, {Dvorak}, {Speith}, \&
  {Sch{\"a}fer}}]{maidvo13b}
{Maindl}, T.~I., {Dvorak}, R., {Speith}, R., \& {Sch{\"a}fer}, C. 2014, ArXiv
  e-print arXiv:1401.0045

\bibitem[{{Maindl} {et~al.}(2013){Maindl}, {Sch{\"a}fer}, {Speith}, {S{\"u}li},
  {Forg{\'a}cs-Dajka}, \& {Dvorak}}]{maisch13}
{Maindl}, T.~I., {Sch{\"a}fer}, C., {Speith}, R., {et~al.} 2013, Astronomische
  Nachrichten, 334, 996

\bibitem[{{Marcq}(2012)}]{Marcq2012}
{Marcq}, E. 2012, Journal of Geophysical Research (Planets), 117, 1001

\bibitem[{{Monaghan} \& {Gingold}(1983)}]{mongin83}
{Monaghan}, J.~J. \& {Gingold}, R.~A. 1983, Journal of Computational Physics,
  52, 374

\bibitem[{{Morbidelli} {et~al.}(2012){Morbidelli}, {Lunine}, {O'Brien},
  {Raymond}, \& {Walsh}}]{Morbidelli-et-al-2012}
{Morbidelli}, A., {Lunine}, J.~I., {O'Brien}, D.~P., {Raymond}, S.~N., \&
  {Walsh}, K.~J. 2012, Annual Review of Earth and Planetary Sciences, 40, 251

\bibitem[{{Ribas} {et~al.}(2005){Ribas}, {Guinan}, {G{\"u}del}, \&
  {Audard}}]{Ribas-et-al-2005}
{Ribas}, I., {Guinan}, E.~F., {G{\"u}del}, M., \& {Audard}, M. 2005, \apj, 622,
  680

\bibitem[{{Sch{\"a}fer}(2005)}]{sch05}
{Sch{\"a}fer}, C. 2005, Dissertation, Eberhard-Karls-Universit{\"a}t
  T{\"u}bingen

\bibitem[{{Sch{\"a}fer} {et~al.}(2007){Sch{\"a}fer}, {Speith}, \&
  {Kley}}]{schspe07}
{Sch{\"a}fer}, C., {Speith}, R., \& {Kley}, W. 2007, \aap, 470, 733

\bibitem[{{Shematovic} \& {et al.}(2014)}]{Shematovich-et-al-2014}
{Shematovic}, V.~I. \& {et al.} 2014, submitted

\bibitem[{{Tian} {et~al.}(2009){Tian}, {Kasting}, \& {Solomon}}]{Tian2009}
{Tian}, F., {Kasting}, J.~F., \& {Solomon}, S.~C. 2009, \grl, 36, 2205

\bibitem[{{Tian} {et~al.}(2005){Tian}, {Toon}, {Pavlov}, \& {De
  Sterck}}]{Tian2005}
{Tian}, F., {Toon}, O.~B., {Pavlov}, A.~A., \& {De Sterck}, H. 2005, Science,
  308, 1014

\bibitem[{{Tillotson}(1962)}]{til62}
{Tillotson}, J.~H. 1962, Metallic Equations of State for Hypervelocity Impact,
  Tech. Rep. General Atomic Report GA-3216, General Dynamics, San Diego, CA

\bibitem[{{Urey}(1955)}]{Urey-1955}
{Urey}, H.~C. 1955, Proceedings of the National Academy of Science, 41, 127

\bibitem[{{Weidenschilling}(1977)}]{wei77}
{Weidenschilling}, S.~J. 1977, \apss, 51, 153

\bibitem[{Wyatt(2009)}]{wya09}
Wyatt, M. 2009, in Lecture Notes in Physics, Vol. 758, Small Bodies in
  Planetary Systems, ed. I.~Mann, A.~Nakamura, \& T.~Mukai (Springer Berlin
  Heidelberg), 1--34

\bibitem[{{Yelle}(2004)}]{Yelle2004}
{Yelle}, R.~V. 2004, \icarus, 170, 167

\bibitem[{{Zahnle} {et~al.}(1990){Zahnle}, {Kasting}, \&
  {Pollack}}]{Zahnle-et-al-1990}
{Zahnle}, K., {Kasting}, J.~F., \& {Pollack}, J.~B. 1990, \icarus, 84, 502

\end{thebibliography}

\end{document}